\documentclass[]{spie}

\usepackage{amsmath,amsfonts,amssymb}
\usepackage{graphicx}
\usepackage[colorlinks=true, allcolors=blue]{hyperref}
\usepackage{booktabs}
\usepackage{multirow}
\usepackage[table]{xcolor}
\usepackage{enumitem}
\usepackage{caption}
\usepackage{capt-of}
\definecolor{lightgreen}{RGB}{242,250,242}
\title{MedTimeQ: Query-guided fusion of time-series foundation models and LLMs for multi-lead ECG classification}

\author[a]{Kiana Taktehrani}
\author[a]{Maryam Hosseini}
\author[a]{Arash Mohammadi}
\affil[a]{Cybersecurity and Intelligent Systems Engineering, Concordia University, Montreal, Canada}

\begin{document} 
\maketitle

\vspace{-.1in}
\begin{abstract}
Accurate classification of medical time-series, such as Electrocardiogram (ECG), requires models that capture subtle waveform morphology, long-range temporal structure, inter-lead dependencies, and the clinical semantics needed to distinguish related cardiac conditions. Existing Large Language Model (LLM)-based approaches often reprogram physiological signal patches directly into the language embedding space. This strategy can underuse complementary knowledge learned by pretrained time-series models and transfer redundant or diagnostically irrelevant information to the LLM. We propose \textbf{MedTimeQ}, a parameter-efficient framework that integrates a lead-aware temporal foundation encoder with an LLM-reprogrammed ECG pathway through gated residual fusion. The MedTimeQ Query Former uses learned diagnostic queries to select and compress the fused representation into a compact set of informative soft tokens for contextualization by a frozen LLM. Semantic prototype learning further guides classification using multiple clinical descriptions for each diagnostic category, statistics-based prompt selection, and knowledge distillation. Both foundation backbones remain frozen, limiting training to lightweight interface and classification components. We evaluate MedTimeQ on PTB–XL, Chapman–Shaoxing, and MIT–BIH for recording-level diagnosis, rhythm classification, and heartbeat classification, respectively. MedTimeQ achieves the best performance in 11 of 12 dataset–metric comparisons against representative time-series and LLM-based methods, with consistent gains in class-balanced metrics across all three tasks. On PTB–XL, it achieves 76.93\% accuracy and 58.14\% macro-F1, surpassing MedTsLLM by 10.08 and 7.07 percentage points, respectively. These results suggest that selectively integrating pretrained temporal representations with language-based clinical semantics improves ECG classification over direct LLM reprogramming alone. The code is publicly available at \url{https://github.com/kkianamm/MedTimeQ}.
\end{abstract}


\section{INTRODUCTION} \label{sec:intro}
Medical time-series data, such as electrocardiograms (ECGs), electroencephalograms (EEGs), respiratory waveforms, and continuously monitored vital signs, play a fundamental role in disease screening, clinical monitoring, and computer-aided diagnosis \cite{ansari2025survey}. ECG classification is particularly challenging as diagnostically relevant patterns involve subtle waveform morphology, long-range temporal dependencies, inter-lead relationships, and clinically meaningful label semantics. ECG manifestations of the same condition can vary across patients, devices, and acquisition settings, whereas related conditions may differ only in fine waveform details~\cite{liu2024generalized,al2026benchmarking,al2026pretraining,he2026medmamba}. 
Foundation models provide two complementary sources of prior knowledge. Time-series foundation models learn transferable physiological representations through large-scale pretraining~\cite{mckeen2025ecg,shu2025clef,nguyen2025tolerantecg,li2026mira,dai2025climb}. In parallel, language-based methods align physiological signals with clinical text or LLM embedding spaces for classification, representation learning, and reasoning. Representative approaches include MedTsLLM, Time2Lang, OpenTSLM, MEIT, GEM, ECG-LLM, K-MERL, SuPreME, and FG-CLEP~\cite{chan2024medtsllm,pillai2025time2lang,langer2025opentslm,wan2025meit,lan2026gem,selivanov2026ecg,liu2025knowledge,cai2025supreme,li2025fine}. Clinically grounded prompt ensembles can further improve biomedical semantic adaptation~\cite{koleilat2025biomedcoop}. Despite these advances, existing approaches do not jointly integrate lead-aware temporal representations, adaptive fusion of temporal and LLM-compatible features, learned-query signal-to-language transfer, and semantically grounded prototype learning within a unified ECG classification framework. 


To address these limitations, we introduce MedTimeQ, a query-guided framework that first integrates temporal and language-aligned ECG representations and then selectively transfers diagnostically relevant evidence to a frozen LLM. Clinically grounded semantic prototypes define the final decision space. The central premise is that pretrained temporal models preserve physiological morphology and dynamics, whereas LLMs provide high-level semantic structure. MedTimeQ learns a compact interface that exploits these complementary capabilities while keeping both foundation backbones frozen. Although evaluated on ECG data, the formulation is applicable across different ECG label spaces and classification settings, including recording-, rhythm-, and beat-level analysis. The principal contributions of this study are summarized below:
\begin{itemize}[topsep=0pt, noitemsep]
\item \textbf{\textit{Query-Guided Temporal-Language Integration:}} MedTimeQ couples a pretrained time-series foundation model with an LLM-reprogrammed pathway through lead-aware temporal encoding, adaptive gated fusion, and learned diagnostic queries.
\item \textbf{\textit{Clinically Grounded Prediction:}} We introduce the MQ-Former module, which efficiently transfers fused physiological representations to a frozen LLM. Statistics-filtered class descriptions and knowledge distillation guide classification within a semantic prototype space, explicitly connecting ECG representations to clinical label meaning.
\item \textbf{Parameter-Efficient Multi-Task Validation:} Both foundation backbones remain frozen, and experiments on three public ECG benchmarks demonstrate consistent gains across recording-level diagnosis, rhythm classification, and heartbeat classification.
\end{itemize}
%
\begin{figure*}[t]
\centering
\includegraphics[width=0.75\textwidth]{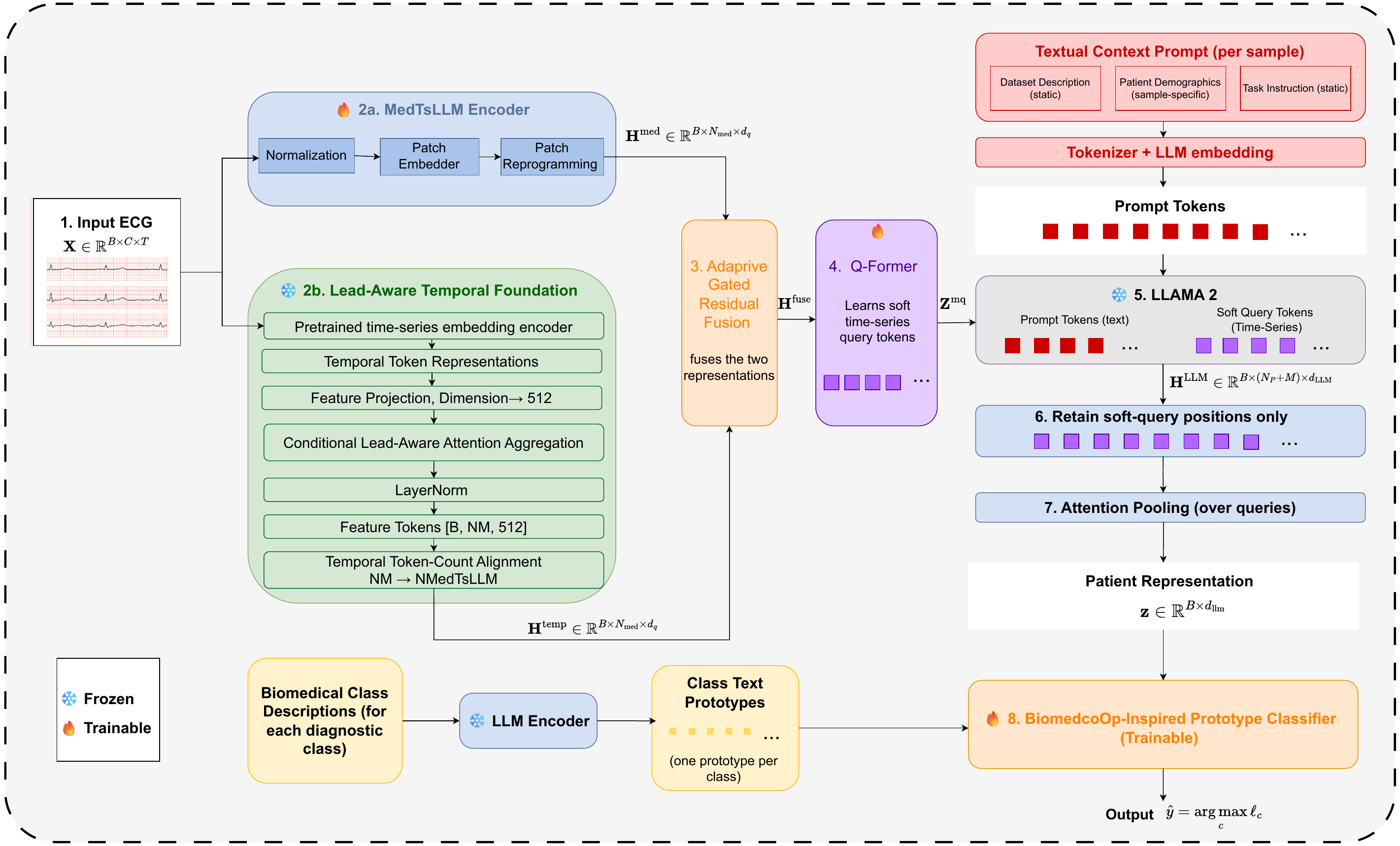}
\vspace{0.025in}
\caption{\footnotesize Overview of MedTimeQ. The MedTsLLM pathway and the proposed lead-aware temporal foundation module are aligned through gated token-level fusion. MQ-Former selectively compresses the fused ECG memory into soft tokens for LLM contextualization, and clinically grounded semantic prototypes define the final decision space.}
\label{fig:main_architecture}
\vspace{-0.15in}
\end{figure*}

\section{The MedTimeQ Framework} \label{sec:method}
\noindent\textbf{Framework Overview:}
Let $\mathbf{X}\in\mathbb{R}^{T\times C}$ denote a multichannel ECG, $\mathbf{P}$ an optional clinical prompt, and $y\in\{1,\ldots,K\}$ its label. MedTimeQ preserves time-series inductive bias, transfers only diagnostically relevant evidence to the LLM, and predicts in a clinically meaningful semantic space. As shown in Fig.~\ref{fig:main_architecture}, an LLM-reprogrammed encoder and a lead-aware temporal foundation encoder produce complementary token streams. Gated fusion combines them, MQ-Former maps the fused memory to a compact query sequence, and the frozen LLM contextualizes these tokens before semantic prototype classification.

\noindent\textbf{LLM-Reprogrammed Signal Encoder:}
The first branch follows MedTsLLM~\cite{chan2024medtsllm}. After normalization and temporal patching, projected ECG patches query a reduced set of embeddings derived from the frozen LLM vocabulary. The resulting tokens $\mathbf{H}^{\mathrm{med}}$ retain patch-level signal information in an LLM-compatible space and serve as the semantic bridge between ECG measurements and language tokens.

\noindent\textbf{Lead-Aware Temporal Foundation Encoder:}
The second branch supplies temporal knowledge that is not naturally encoded by an LLM. A frozen MOMENT-1-base backbone~\cite{goswami2024moment} processes the channel-first ECG, and trainable projections map its outputs to the shared dimension. Learned lead embeddings preserve channel identity, while attention aggregates complementary evidence. For temporal token $n$ and lead $c$,
\begin{equation}
a_{c,n}=\frac{\exp\!\left(\mathbf{w}_{l}^{\top}\tanh(\mathbf{W}_{l}\mathbf{h}_{c,n}+\mathbf{r}_{c})\right)}
{\sum_{c'=1}^{C}\exp\!\left(\mathbf{w}_{l}^{\top}\tanh(\mathbf{W}_{l}\mathbf{h}_{c',n}+\mathbf{r}_{c'})\right)},
\qquad
\widehat{\mathbf{h}}_{n}^{\mathrm{temp}}=\sum_{c=1}^{C}a_{c,n}\mathbf{h}_{c,n},
\label{eq:lead_attention}
\end{equation}
where $\mathbf{r}_{c}$ is the embedding of lead $c$. This lead-aware pooling retains morphology and cross-lead structure that are typically underrepresented in direct language reprogramming.

\noindent\textbf{Query-Guided Temporal-Language Integration:}
The two token streams are projected to dimension $d_q$, and the temporal sequence is interpolated to the MedTsLLM token length. For aligned tokens $\widetilde{\mathbf{h}}_{n}^{\mathrm{med}}$ and $\widetilde{\mathbf{h}}_{n}^{\mathrm{temp}}$, let $\mathbf{u}_{n}$ denote their concatenation. A feature-wise gate controls the temporal residual as follows
\begin{equation}
\mathbf{g}_{n}=\sigma\!\left(\mathbf{W}_{g2}\phi(\mathbf{W}_{g1}\mathbf{u}_{n})\right),\quad
\Delta\mathbf{h}_{n}=\mathbf{W}_{d2}\phi(\mathbf{W}_{d1}\mathbf{u}_{n}),\quad
\mathbf{h}_{n}^{\mathrm{fuse}}=\operatorname{LN}\!\left(\widetilde{\mathbf{h}}_{n}^{\mathrm{med}}+\mathbf{g}_{n}\odot\Delta\mathbf{h}_{n}\right),
\label{eq:fusion}
\end{equation}
where $\phi$ is GELU. The residual retains the LLM-compatible representation, while the gate determines where and how strongly temporal evidence modifies it. Passing every fused token to the LLM would be redundant and expensive. MQ-Former instead maintains $M$ learned diagnostic queries. Each layer applies query self-attention followed by cross-attention to the fused ECG memory as follows
\begin{equation}
\widehat{\mathbf{Q}}^{(\ell)}=\operatorname{LN}\!\left(\mathbf{Q}^{(\ell-1)}+\operatorname{MHSA}(\mathbf{Q}^{(\ell-1)})\right),\quad
\mathbf{Q}^{(\ell)}=\operatorname{LN}\!\left(\widehat{\mathbf{Q}}^{(\ell)}+\operatorname{MHCA}(\widehat{\mathbf{Q}}^{(\ell)},\mathbf{H}^{\mathrm{fuse}},\mathbf{H}^{\mathrm{fuse}})\right).
\label{eq:mqformer}
\end{equation}
After $L_q$ layers, a learned projection maps the queries to the LLM embedding dimension. Combined with prompt embeddings, these fixed-length soft tokens allow individual queries to specialize in complementary morphological, temporal, or cross-lead evidence.

\noindent\textbf{Clinically Grounded Semantic Prediction:} For each class, multiple clinical descriptions encode characteristic morphology and diagnostic criteria. The frozen language model embeds these descriptions; statistics-based selection removes poorly aligned prompts, and the retained embeddings are averaged and normalized to form class prototype $\overline{\mathbf{t}}_c$. The normalized ECG representation $\widehat{\mathbf{s}}$ is classified by cosine similarity, $z_c=\exp(\eta)\widehat{\mathbf{s}}^{\top}\overline{\mathbf{t}}_c$, with learnable log-temperature $\eta$. Selected prototypes also serve as a teacher for the trainable prediction head.

\noindent\textbf{Learning Objective \& Adaptation:} The training objective combines cross-entropy with auxiliary losses on the MedTsLLM, temporal, and MQ-Former representations; cross-branch alignment; query-diversity regularization; and semantic distillation given by
\begin{equation}
\mathcal{L}=\mathcal{L}_{\mathrm{cls}}+\lambda_{\mathrm{med}}\mathcal{L}_{\mathrm{med}}+\lambda_{\mathrm{temp}}\mathcal{L}_{\mathrm{temp}}+\lambda_{\mathrm{mq}}\mathcal{L}_{\mathrm{mq}}+\lambda_{\mathrm{align}}\mathcal{L}_{\mathrm{align}}+\lambda_{\mathrm{div}}\mathcal{L}_{\mathrm{div}}+\lambda_{\mathrm{KD}}\mathcal{L}_{\mathrm{KD}}.
\label{eq:objective}
\end{equation}
Training is limited to the MedTsLLM adaptation layers, temporal projections, lead aggregation, fusion, MQ-Former, pooling, auxiliary heads, and semantic alignment projection.

\begin{table*}[t]
\centering
\caption{\footnotesize Classification and ablation results (\%). Panels (a)-(c) compare
 methods on three ECG benchmarks; panel (d) reports
component-wise ablations on PTB-XL. The best and second-best comparison
results are shown in bold and underlined, respectively.}
\label{tab:combined_results}

\renewcommand{\arraystretch}{1.18}
\setlength{\tabcolsep}{5.5pt}
\footnotesize

\begin{minipage}[t]{0.485\textwidth}
\centering
\textbf{(a) PTB--XL classification}\\[3pt]

\begin{tabular}{lcccc}
\toprule
\textbf{Method} & \textbf{Acc.} & \textbf{F1} &
\textbf{Prec.} & \textbf{Rec.} \\
\midrule
GPT4TS~\cite{zhou2023one}
& 57.58 & 40.83 & 48.23 & 40.68 \\

PatchTST~\cite{nie2022time}
& 63.47 & 49.19 & 52.12 & 47.78 \\

OpenTSLM
& 65.79 & 41.52 & 44.61 & 40.55 \\

MedTsLLM
& 66.85 & 51.07 & \underline{54.20} & \underline{49.85} \\

GEM
& \underline{69.88} & \underline{51.32} & 54.18 & 49.49 \\

\rowcolor{lightgreen}
\textbf{MedTimeQ}
& \textbf{76.93} & \textbf{58.14}
& \textbf{60.25} & \textbf{57.70} \\
\bottomrule
\end{tabular}
\end{minipage}
\hfill
\begin{minipage}[t]{0.485\textwidth}
\centering
\textbf{(b) Chapman--Shaoxing classification}\\[3pt]

\begin{tabular}{lcccc}
\toprule
\textbf{Method} & \textbf{Acc.} & \textbf{F1} &
\textbf{Prec.} & \textbf{Rec.} \\
\midrule
GPT4TS
& 75.60 & 33.97 & 38.46 & 35.45 \\

PatchTST
& 75.20 & 34.95 & 47.38 & 35.82 \\

OpenTSLM
& 83.62 & 48.78 & 54.00 & 48.36 \\

MedTsLLM
& \underline{86.08} & \underline{54.31}
& \underline{56.51} & \underline{53.38} \\

GEM
& 84.97 & 51.56 & 53.76 & 52.06 \\

\rowcolor{lightgreen}
\textbf{MedTimeQ}
& \textbf{91.02} & \textbf{60.84}
& \textbf{69.49} & \textbf{61.66} \\
\bottomrule
\end{tabular}
\end{minipage}

\vspace{7pt}

\begin{minipage}[t]{0.485\textwidth}
\centering
\textbf{(c) MIT--BIH classification}\\[3pt]

\begin{tabular}{lcccc}
\toprule
\textbf{Method} & \textbf{Acc.} & \textbf{F1} &
\textbf{Prec.} & \textbf{Rec.} \\
\midrule
GPT4TS
& 84.51 & \underline{35.42}
& 34.39 & \underline{40.34} \\

PatchTST
& \underline{89.81} & 26.22
& \underline{38.36} & 24.96 \\

OpenTSLM
& 86.35 & 34.71 & 33.43 & 36.83 \\

MedTsLLM
& 89.78 & 34.33 & 34.07 & 36.42 \\

GEM
& 86.68 & 34.39 & 31.49 & \textbf{40.95} \\

\rowcolor{lightgreen}
\textbf{MedTimeQ}
& \textbf{93.28} & \textbf{38.41}
& \textbf{40.58} & 37.57 \\
\bottomrule
\end{tabular}
\end{minipage}
\hfill
\begin{minipage}[t]{0.485\textwidth}
\centering
\textbf{(d) Component-wise ablation on PTB--XL}\\[3pt]

\begin{tabular}{lcccc}
\toprule
\textbf{Configuration} & \textbf{Acc.} & \textbf{F1} &
\textbf{Prec.} & \textbf{Rec.} \\
\midrule
MedTsLLM
& 66.85 & 51.07 & 54.20 & 49.85 \\

+LTFM
& 68.83 & 57.47 & 59.74 & 56.64 \\

+SPL
& 67.73 & 53.70 & 58.40 & 54.34 \\

+MQF
& 65.78 & 53.21 & 56.94 & 52.99 \\

\rowcolor{lightgreen}
\textbf{Full MedTimeQ}
& \textbf{76.93} & \textbf{58.14}
& \textbf{60.25} & \textbf{57.70} \\
\bottomrule
\end{tabular}

\end{minipage}
\vspace{-.15in}
\end{table*}

\vspace{-.1in}
\section{Experiments} \label{sec:experiments}

\noindent\textbf{Experimental Setup:} The two representation streams use a shared 512-dimensional space. MQ-Former contains four layers, 32 learned queries, and eight attention heads; its outputs are projected to the frozen Llama-2-7B embedding space~\cite{touvron2023llama}. Models are trained for 20 epochs with Adam, batch size 16, learning rate $10^{-4}$, zero weight decay, dropout 0.1, and bfloat16 precision. The best validation checkpoint is evaluated. Implementations use PyTorch, Transformers 4.57.6, Python 3.10.12, CUDA 13.2, and an NVIDIA H100 80GB GPU.
All inputs contain 512 samples. \textbf{PTB-XL} contains 21,799 ten-second, 12-lead ECGs from 18,869 patients grouped into five diagnostic superclasses~\cite{wagner2020ptb,PhysioNet-ptb-xl-1.0.3,goldberger2000physiobank}. We use folds 1--8/9/10 for training/validation/testing. \textbf{MIT--BIH} contains 48 thirty-minute recordings from 47 patients~\cite{moody2001impact}. We use the 44 non-pacemaker recordings and the inter-patient DS1/DS2 protocol, reserve records 114, 119, 208, and 223 for validation, and classify beat-centered windows into the five AAMI classes. \textbf{Chapman--Shaoxing} contains 45,152 ten-second, 12-lead recordings at 500~Hz~\cite{PhysioNet-ecg-arrhythmia-1.0.0}. We retain unambiguous single-label recordings from nine rhythm classes, resample each record to 512 points, normalize with training-set lead statistics, and use stratified 80/10/10 splits. We report accuracy and macro-averaged F1, precision, and recall; macro-F1 is the primary metric because of class imbalance.

\noindent\textbf{Does temporal-language integration improve ECG classification?}
Table~\ref{tab:combined_results}(a)--(c) compares MedTimeQ with
representative methods across the three ECG benchmarks. It shows that MedTimeQ ranks first in 11 of 12 dataset--metric comparisons. On PTB--XL, it reaches 76.93\% accuracy and 58.14\% macro-F1, exceeding the strongest competing results by 7.05 and 6.82 points. Relative to MedTsLLM, the gains are 10.08 accuracy points and 7.07 macro-F1 points. On Chapman--Shaoxing, MedTimeQ improves on MedTsLLM by 4.94 points in accuracy, 6.53 in macro-F1, 12.98 in macro-precision, and 8.28 in macro-recall. On MIT--BIH, it obtains the highest accuracy, macro-F1, and macro-precision; GEM attains higher macro-recall. The high accuracy but lower macro-F1 on MIT--BIH is consistent with its severe class imbalance. The gains span recording-level diagnosis, rhythm classification, and heartbeat classification, despite differences in acquisition protocols and label spaces. This consistency supports the framework's central thesis: pretrained temporal structure and language-based semantics provide complementary information, and selective query-guided transfer can exploit both.

\noindent\textbf{What contributes to the gains?} The ablation in Table~\ref{tab:combined_results}(d) shows that adding the lead-aware temporal encoder with fusion to MedTsLLM provides the largest individual increase in macro-F1 and recall, from 51.07\% to 57.47\% and from 49.85\% to 56.64\%. This indicates that pretrained temporal and cross-lead information addresses limitations of the reprogrammed pathway. Semantic prototype learning improves all class-balanced metrics, consistent with better discrimination among related diagnoses. MQ-Former alone lowers accuracy slightly but improves macro-F1, precision, and recall, suggesting a more balanced decision boundary. The full model yields the highest accuracy and precision and the best overall metric profile, showing that temporal encoding, selective query transfer, and semantic supervision are complementary rather than interchangeable.

\noindent\textbf{Limitations:} The study is retrospective and limited to public ECG benchmarks; modality-agnostic applicability has not yet been established experimentally. Inference cost also depends on the selected foundation backbones. Prospective and external clinical validation, evaluation under missing or corrupted leads, and replacement of the current backbones with smaller models are important next steps.

\section{Conclusion}
We introduced MedTimeQ, a query-guided framework that couples pretrained temporal and language foundation models for ECG classification. A lead-aware temporal encoder and gated fusion preserve physiological structure, MQ-Former selectively transfers diagnostic evidence to a frozen LLM, and clinical text prototypes ground the prediction space. Across three public benchmarks, this parameter-efficient design consistently improves on representative time-series and LLM-based methods. These results indicate that temporal and language foundation models are complementary sources of knowledge whose interaction can be learned through compact query interfaces. Future work will assess whether this paradigm extends to other modalities and clinical~tasks.

\bibliographystyle{spiebib}
\bibliography{report} 
\end{document}